\documentstyle[12pt]{article}
\begin{document}
\begin{center}
{\bf{INFRARED REGION OF QCD AND CONFINEMENT}}
\end{center}

\vspace{1.0cm}

\begin{center}
R.Parthasarathy{\footnote{e-mail address:
sarathy@imsc.ernet.in}} \\
The Institute of Mathematical Sciences \\
C.P.T.Campus, Taramani Post \\
Chennai - 600 113 , India \\
\end{center}

\vspace{4.0cm}

{\noindent{\it{\bf{Abstract}}}}

\vspace{0.5cm}

Gauge field configurations appropriate for the infrared region 
of QCD is proposed. Using the usual QCD action, confinement is
realized as in the London theory of Meissner effect. 

\vspace{0.5cm}


\newpage

\vspace{0.5cm}

The mechanism of confinement of gluons and quarks is very
important to understand. A suggestion of 'tHooft$^{1}$ and
Mandelstam$^{2}$ is that magnetic monopoles must play a crucial
role in confinement. The high energy limit of QCD has been
studied by perturbation theory due to the small coupling
strength of the QCD running coupling constant and experimental
results confirm the use of perturbation theory in this
region$^{3}$. Here the gluons are massless and the color
symmetry is an exact symmetry. At low energies, the coupling
is strong and non-perturbative methods are to be invoked. This
region is not completely understood. Color confinement is
possibly due to a dual Meissner effect$^{2,4,5}$ and this
suggests that in the infrared region, some other variables,
other than $A^a_{\mu}$, may be more relevant.  
The idea of 'tHooft$^{1}$ is to make an
Abelian projection and Kondo$^{6}$ has examined this for
$SU(2)$ gauge theory by integrating the non-Abelian degrees of
freedom to obtain an Abelian projected effective field theory.
Here, the relation between the non-Abelian gauge fields and
the monopole configuration is not clear. 
Recently, Faddeev and Niemi$^{7}$ have proposed a set of
variables for describing the infra-red limit of 4-dimensional
$SU(2)$ and $SU(N)$ gauge theory. 
This is a continuation of the  earlier
works of Corrigan et.al.,$^{8}$ and Cho$^{9}$.  
In Ref.9,  a complex scalar field $\phi$ for
the monopole has been introduced and confinement is suggested
by dynamically breaking (Coleman-Weinberg mechanism) the
magnetic symmetry. On the other hand, Faddeev and Niemi$^{7}$
propose a non-linear sigma model action as relevant in the
infra-red limit of Yang-Mills theory, motivated by  Wilsonian
renormalization group arguments.

\vspace{1.0cm}

In this Letter, a gauge field configuration for the infra-red
limit of QCD is proposed in which the magnetic confinement of
gluons and quarks will be realized. By considering the dual 
version, the
electric confinement will as well be realized. 

\vspace{1.0cm}

The proposed $SU(3)$ gauge field configurations $A^a_{\mu}$  
, for the infrared region of QCD, is taken to
satisfy
\begin{eqnarray}
D^{ab}_{\mu}{\omega}^b\ =\ {\partial}_{\mu}{\omega}^a + g
f^{acb}A^c_{\mu}{\omega}^b &=& 0,
\end{eqnarray}
where ${\omega}^a\ \in \ su(3)$ and are chosen such that
\begin{eqnarray}
{\omega}^a{\omega}^a &=& 1 \nonumber \\
d^{abc}{\omega}^b{\omega}^c& = & 
\frac{1}{\sqrt{3}}\ {\omega}^a,
\end{eqnarray}
with $d^{abc}$ as the symmetric Gell-Mann $SU(3)$ tensor. 
Now, in the infra-red region, the gauge group manifold is
not the complete $su(3)$ group manifold spanned by
arbitrary ${\omega}^a$'s but a {\it{submanifold spanned by
those ${\omega}^a$'s satisfying (2) }} and the relevant gauge
field configurations $A^a_{\mu}$ are determined by (1).
Classically, the Wu and Wu$^{10}$ ansatz for the $SU(3)$ gauge
field, satisfying the Euler-Lagrange equations of pure $SU(3)$
gauge theory and the construction  ${\omega}^a\ =\
d^{abc}F^b_{\mu\nu}F^c_{\mu\nu}$ satisfy (1) and (2)$^{11}$. 
{\it{Instead, here we
will choose to solve (1) for $A^a_{\mu}$}}. The second relation
in (2) is very {\it{special}} to $SU(3)$ and will be  
{\it{crucial}} 
in what follows.

\vspace{1.0cm}

A solution to (1)$^{12}$ is
\begin{eqnarray}
A^a_{\mu} &=& C_{\mu}\ {\omega}^a - \frac{4}{3g}f^{abc}\
{\omega}^b\ {\partial}_{\mu}{\omega}^c,
\end{eqnarray}
where $C_{\mu}$ is arbitrary. In obtaining (3), we have
derived a useful relation for ${\omega}^a$'s in (2),
\begin{eqnarray}
\frac{4}{3}f^{abc}f^{edc}{\omega}^b{\omega}^e{\partial}_{\mu}
{\omega}^d &=& - {\partial}_{\mu}{\omega}^a,
\end{eqnarray}
which will be used subsequently. The gauge field configuration
(3) and the submanifold determined by (2) are proposed to
describe the infra-red region of QCD. Relations (1), (2) and
(3) are self-consistent. It is to be noted here that
${\omega}^a\ A^a_{\mu}\ =\ C_{\mu}\ \neq \ 0$, in contrast to
the classical configuration described in Ref.11.  

\vspace{1.0cm}

There are two important consequences that follow from (1).
Usually a mass term for the gauge fields, $m^2
A^a_{\mu}A^a_{\mu}$ cannot be present in the Lagrangian
density due to its gauge non-invariance. Since under a gauge
transformation, the field changes as $\delta A^a_{\mu}\ =\
D^{ab}_{\mu}{\omega}^b$, in view of (1), if we consider gauge
transformations within the submanifold, then such a mass term
is allowed in it. This is similar to defining the state vector
in the Heisenberg picture in quantum mechanics$^{13}$.
${\psi}_H\ =\ exp(-i\frac{Ht}{\hbar}){\psi}_S(t)$, i.e., as
${\psi}_S(t)$ evolves in time, it is brought back to
${\psi}_H\ =\ {\psi}_S(0)$ at every instant. Second, if we
choose a Lorentz covariant gauge ${\partial}_{\mu}A^a_{\mu}\
=\ 0$, for $A^a_{\mu}$'s in (3), then there will be no Gribov
ambiguity$^{14}$ in fixing the gauge in view of (1), 
as long as we are within
the submanifold.

\vspace{1.0cm}

The field strength $F^a_{\mu\nu}\ =\
{\partial}_{\mu}A^a_{\nu}-{\partial}_{\nu}A^a_{\mu}+gf^{abc}
A^b_{\mu}A^c_{\nu}$, for $A^a_{\mu}$ in (3) is calculated,
using the relations (2), (4) and the relations among the $f$
and $d$ tensors$^{15}$, as 
\begin{eqnarray}
F^a_{\mu\nu} &=&
({\partial}_{\mu}C_{\nu}-{\partial}_{\nu}C_{\mu})\ {\omega}^a
 -
\frac{4}{3g}\ f^{abc}{\partial}_{\mu}{\omega}^b{\partial}_
{\nu}{\omega}^c.
\end{eqnarray} 
It is found that $F^a_{\mu\nu}$ is "$SU(3)$ orthogonal" to
${\omega}^a$ $^{16}$, i.e.,
\begin{eqnarray}
f^{abc}\ {\omega}^b\ F^c_{\mu\nu} &=& 0,
\end{eqnarray}
consistent with (1). 
Unlike the case of $SU(2)$ $^{8,9}$, this does not mean that
$F^a_{\mu\nu}$ is along ${\omega}^a$. Although the first term
in (5) is along ${\omega}^a$, the second term is not. If a
tensor field strength $G_{\mu\nu}$ is constructed as
$G_{\mu\nu}\ =\ {\omega}^a F^a_{\mu\nu}$, then $G_{\mu\nu}\
=\ f_{\mu\nu} -
\frac{4}{3g}f^{abc}{\omega}^a{\partial}_{\mu}{\omega}^b
{\partial}_{\nu}{\omega}^c$, where $f_{\mu\nu}\ =\
{\partial}_{\mu}C_{\nu}-{\partial}_{\nu}C_{\mu}$. If  an
action $-\frac{1}{4}\int G^2_{\mu\nu}\ d^4x$ is considered,
then the action becomes,
\begin{eqnarray}
S &=& -\frac{1}{4}\int \{
f^2_{\mu\nu}-\frac{8}{3g}f_{\mu\nu}X_{\mu\nu} + \frac{16}
{9g^2}X_{\mu\nu}X_{\mu\nu}\} d^4x, \nonumber
\end{eqnarray}
where $X_{\mu\nu}\ =\
f^{abc}{\omega}^a{\partial}_{\mu}{\omega}^b{\partial}_{\nu}
{\omega}^c$. This action coincides with the structure of the
effective action for Abelian projected QCD of Kondo$^6$.
Instead of this procedure, we will consider the {\it{usual}} 
action for pure QCD.
\begin{eqnarray}
I &=& -\frac{1}{4}\int (F^a_{\mu\nu})^2\ d^4x, \nonumber \\
 &=& -\frac{1}{4}\int \{ f^2_{\mu\nu} -
\frac{8}{3g}f_{\mu\nu}(f^{abc}{\omega}^a{\partial}_{\mu}
{\omega}^b{\partial}_{\nu}{\omega}^c) \nonumber \\
&+& \frac{16}{9g^2}f^{abc}f^{aed}{\partial}_{\mu}
{\omega}^b{\partial}_{\nu}{\omega}^c{\partial}_{\mu}
{\omega}^e{\partial}_{\nu}{\omega}^d\ \} d^4x.
\end{eqnarray} 
This action has $U(1)$ symmetry, $C_{\mu}\ \rightarrow \
C_{\mu} + {\partial}_{\mu}\Lambda$. It will be convenient to
rescale the ${\omega}^a$'s as $g^{\frac{1}{3}}{\omega}^a$ and
then in the strong coupling limit (The last term in (7) is
omitted hereafter in the strong coupling limit. It can be
included at the end and this will not change the conclusions), 
the above action becomes
\begin{eqnarray}
I &\simeq & -\frac{1}{4}\int \{ f^2_{\mu\nu} -
\frac{8}{3}f_{\mu\nu}(f^{abc}{\omega}^a{\partial}_{\mu}
{\omega}^b{\partial}_{\nu}{\omega}^c)\} d^4x,
\end{eqnarray}
showing the Abelian dominance for
the infra-red limit of QCD. Here the field strength
$f_{\mu\nu}$ is coupled to $X_{\mu\nu}\ =\
f^{abc}{\omega}^a{\partial}_{\mu}{\omega}^b{\partial}_{\nu}
{\omega}^c$. First of all it is seen that
${\partial}_{\mu}X_{\mu\nu}\ =\ f^{abc}{\omega}^a {\partial}
_{\mu}({\partial}_{\mu}{\omega}^b{\partial}_{\nu}{\omega}^c)
\ \neq \ 0$ and the dual strength $\tilde{X}_{\mu\nu}\ =\ 
\frac{1}{2}{\epsilon}_{\mu\nu\alpha\beta}X_{\alpha\beta}$
violates the Bianchi identity,
\begin{eqnarray}
{\partial}_{\mu}\tilde{X}_{\mu\nu}&=& -\frac{2}{3}{\epsilon}
_{\mu\nu\alpha\beta}f^{abc}{\partial}_{\mu}{\omega}^a
{\partial}_{\alpha}{\omega}^b{\partial}_{\beta}   
{\omega}^c \ \neq \ 0.
\end{eqnarray}
This along with
\begin{eqnarray}
-\frac{2}{3}\oint_{S}{\epsilon}_{\mu\nu\alpha\beta}f^{abc}
{\omega}^a{\partial}_{\alpha}{\omega}^b{\partial}_{\beta}
{\omega}^c dx^{\mu}\wedge dx^{\nu}\ =\ -\frac{4}{3}\oint_
{S}\tilde{X}_{\mu\nu}dx^{\mu}\wedge dx^{\nu} \ =\ 4\pi M,
\end{eqnarray}
a topological invariant, imply that magnetic monopoles are
present in (8) and coupled to $f_{\mu\nu}$. Second, as in the
infra-red region, we proposed  the gauge field configuration (3)
in the submanifold determined by (2), a mass term $m^2
A^a_{\mu}A^a_{\mu}$ can be introduced. From (3) it follows
that
\begin{eqnarray}
A^a_{\mu}A^a_{\mu}&=&
C_{\mu}C_{\mu}+\frac{4}{3g^2}{\partial}_{\mu}{\omega}^a
{\partial}_{\mu}{\omega}^a.
\end{eqnarray}
After rescaling ${\omega}^a$ by $g^{\frac{1}{3}}$ as used in
(8) and taking the strong coupling limit,
$A^a_{\mu}A^a_{\mu}\ \simeq \ C_{\mu}C_{\mu}$. Thus a mass
term for $A^a_{\mu}$ induces a mass term for $C_{\mu}$. This
breaks the $U(1)$ invariance. A mass term for $C_{\mu}$ can
also be induced by Coleman-Weinberg mechanism by introducing
complex scalars $\phi$ minimally coupled to $C_{\mu}$ as in
the works of Mandelstam$^2$, 'tHooft$^1$ and Cho$^{9}$. Then the
action (8) becomes
\begin{eqnarray}
I&=&-\frac{1}{4}\int \{
f^2_{\mu\nu}-\frac{8}{3}f_{\mu\nu}X_{\mu\nu}+m^2C_{\mu}
C_{\mu}\} d^4x.
\end{eqnarray}
The partition function for the above is
\begin{eqnarray}
Z&=& \int [dC_{\mu}]exp(-S),
\end{eqnarray}
and by integrating over the $C_{\mu}$ field 
(we suppress the
gauge fixing and ghost terms here as we have only Abelian
fields), we find
\begin{eqnarray}
Z&=& det^{-1}(\Box  - \frac{m^2}{2})\
exp(\int \frac{8}{9}{\partial}_{\mu}X_{\mu\nu}(\Box 
  -
 \frac{m^2}{2})^{-1}{\partial}_{\rho}X_{\rho\nu} d^4x ).
\end{eqnarray}
The effective Lagrangian from this is, apart from the constant
divergent factor,
\begin{eqnarray}
L_{eff}&\simeq &-\frac{8}{9} {\partial}_{\mu}X_{\mu\nu} (\Box 
  - \frac{m^2}{2})^{-1}{\partial}_{\rho}X_{\rho
\nu},
\end{eqnarray}
and as ${\partial}_{\mu}X_{\mu\nu}$ is non-vanishing, this
form is identical to the London case of magnetic confinement.
This provides a magnetic confinement of gluons in the
submanifold providing a scenario suggested by Nambu$^{5}$
that color confinement could occur similar to the magnetic
confinement in an ordinary superconductor due to the Meissner
effect.   

\vspace{1.0cm}

It is possible to provide electric confinement by rewriting
the action (8) in its dual form. The generating functional for
(8) is
\begin{eqnarray}
Z&=& \int [dC_{\mu}]exp(\frac{1}{4}\int \{ f^2_{\mu\nu} -
\frac{8}{3}f_{\mu\nu}X_{\mu\nu}\}d^4x.
\end{eqnarray}
Introducing the dual field strength ${\cal{G}}_{\mu\nu}$ dual
to $f_{\mu\nu}$, we can rewrite the above $Z$ as
\begin{eqnarray}
Z&=&\int [dC_{\mu}][d{\cal{G}}_{\mu\nu}]exp{\left(
-\frac{1}{4}{\cal{G}}^2_{\mu\nu}+\frac{1}{2}{\cal{G}}_{\mu\nu}
f_{\mu\nu}-\frac{2}{3}f_{\mu\nu}X_{\mu\nu}\right)}d^4x,
\end{eqnarray}
as the functional integration over ${\cal{G}}_{\mu\nu}$ brings
(15). From (16), variation with respect to $C_{\mu}$-field
($f_{\mu\nu}\ =\
{\partial}_{\mu}C_{\nu}-{\partial}_{\nu}C_{\mu}$) gives
${\partial}_{\mu}\{ {\cal{G}}_{\mu\nu}-\frac{4}{3}X_{\mu\nu}\}\
=\ 0$, which is solved for ${\cal{G}}_{\mu\nu}$ as
\begin{eqnarray}
{\cal{G}}_{\mu\nu} &=&
{\epsilon}_{\mu\nu\lambda\sigma}{\partial}_{\lambda}{\tilde{A
}}_{\sigma}+\frac{4}{3}X_{\mu\nu}.
\end{eqnarray}
The field $\tilde{A}_{\mu}$ serves as dual to $C_{\mu}$. Use
of (17) in (16) eliminates $C_{\mu}$-field and we obtain
\begin{eqnarray}
Z&=& \int
[d\tilde{A}_{\mu}]exp[\frac{1}{4}\int\{({\partial}_{\lambda}
\tilde{A}_{\sigma}-{\partial}_{\sigma}\tilde{A}_{\lambda})^2
\nonumber \\
&+&\frac{8}{3}X_{\mu\nu}{\epsilon}_{\mu\nu\lambda\sigma}
{\partial}_{\lambda}\tilde{A}_{\sigma}+\frac{16}{9}
X_{\mu\nu}X_{\mu\nu}\}d^4x].
\end{eqnarray}
This action possess dual $U(1)$ invariance and coincides with
the Abelian projected effective field theory for QCD based on
$SU(3)$ in its {\it{dual}} form. Denoting
${\tilde{X}}_{\lambda\sigma}\ =\ \frac{1}{2}{\epsilon}_{
\lambda\sigma\mu\nu}X_{\mu\nu}$, we have
\begin{eqnarray}
Z&=& \int [d\tilde{A}_{\mu}] exp[\frac{1}{4}\int \{ (
{\partial}_{\lambda}\tilde{A}_{\sigma}-{\partial}_{\sigma}
\tilde{A}_{\lambda})^2 \nonumber \\
&+& \frac{16}{3}\tilde{X}_{\lambda\sigma}{\partial}_{\lambda}
\tilde{A}_{\sigma}+\frac{16}{9}X_{\mu\nu}X_{\mu\nu}+M^2
\tilde{A}_{\mu}\tilde{A}_{\mu}\}d^4x],
\end{eqnarray}
where a mass term for $\tilde{A}_{\mu}$ is introduced which
can be thought of arising from Coleman-Weinberg mechanism by
invoking complex scalars for example. By functionally
integrating the dual field $\tilde{A}_{\mu}$,  
the effective
Lagrangian density, apart from constant (divergent) factors 
not involving fields, is found to be 
\begin{eqnarray}
{\cal{L}}_{eff}&=&
-\frac{8}{9}{\partial}_{\lambda}\tilde{X}_{\lambda\sigma}
(\Box  - \frac{M^2}{2})^{-1} {\partial}_{\rho}
\tilde{X}_{\rho\sigma} - \frac{4}{9}\tilde{X}_{\mu\nu}
\tilde{X}_{\mu\nu},
\end{eqnarray}
which gives the {\it{electric}} (dual) confinement of gluons
in the submanifold. 

\vspace{1.0cm}

The reason for realizing both magnetic and electric
confinement is the property of $X_{\mu\nu}\ =\
f^{abc}{\omega}^a{\partial}_{\mu}{\omega}^b{\partial}_{\nu}
{\omega}^c$ that both ${\partial}_{\mu}X_{\mu\nu}$ and
${\partial}_{\mu}\tilde{X}_{\mu\nu}$ are non-vanishing. These
fields $X_{\mu\nu}$ and $\tilde{X}_{\mu\nu}$ couple to
$f_{\mu\nu}$ and its dual. While the non-vanishing of
${\partial}_{\mu}\tilde{X}_{\mu\nu}$ violating the Bianchi
identity is attributed to the monopole topological
configuration (in (9) and (10)), that of
${\partial}_{\mu}X_{\mu\nu}$ is due to the construction 
itself. 

\vspace{1.0cm}

Now we consider the QCD action with quarks. The standard
action is 
\begin{eqnarray}
I&=& -\frac{1}{4}\int F^a_{\mu\nu}F^a_{\mu\nu} d^4x + \int
\bar{\psi}i{\gamma}^{\mu}({\partial}_{\mu}+ig\frac{1}{2}
{\lambda}^a\ A^a_{\mu})\psi \ d^4x.
\end{eqnarray}
The first term has been expressed in terms of $C_{\mu}$ and
${\omega}^a$. The Dirac current $j^a_{\mu}\ =\
g\bar{\psi}{\gamma}^{\mu}\frac{1}{2}{\lambda}^a \psi$ couples
to $A^a_{\mu}$ minimally. Using the expression (3) for
$A^a_{\mu}$, we find the action becomes
\begin{eqnarray}
I&=& -\frac{1}{4}\int \{ f^2_{\mu\nu} - \frac{8}{3}f_{\mu\nu}
X_{\mu\nu} + m^2 C_{\mu}C_{\mu}\} d^4x  \nonumber \\
&+& \int \{\bar{\psi}{\gamma}^{\mu}i{\partial}_{\mu}\psi -   
j^a_{\mu}{\omega}^a C_{\mu} -
\frac{4}{3g}j^a_{\mu}f^{abc}{\omega}^b{\partial}_{\mu}
{\omega}^c \} d^4x.
\end{eqnarray} 
In the corresponding partition function, the $C_{\mu}$ field
is integrated to give the effective Lagrangian density
\begin{eqnarray}
L_{eff} &=& -\frac{8}{9}{\partial}_{\mu}X_{\mu\nu}(\Box -
\frac{m^2}{2})^{-1}{\partial}_{\rho}X_{\rho\nu} - \frac{1}{2} 
j^a_{\mu}
{\omega}^a (\Box - \frac{m^2}{2})^{-1}j^b_{\mu}{\omega}^b
\nonumber \\ 
&+& \frac{4}{3}{\partial}_{\mu}X_{\mu\nu}(\Box -
\frac{m^2}{2})^{-1}j^a_{\nu}{\omega}^a - \frac{4}{3g}j^a_{\mu}
f^{abc}{\omega}^b{\partial}_{\mu}{\omega}^c ,
\end{eqnarray}
apart from the kinetic energy term for quarks. This is
identical to the London case of magnetic confinement including
the quarks. In addition the monopole current interacts with
the Dirac current through $(\Box - \frac{m^2}{2})^{-1}$. A
similar procedure can be effected for the dual version by
coupling quarks to dual field $\bar{A}_{\mu}$.

\vspace{1.0cm}

Thus, by considering a submanifold for the group $SU(3)$
defined by both the relations in (2) and proposing the gauge
field $A^a_{\mu}$ configuration (3) satisfying (1), we obtain
a $F^a_{\mu\nu}$ "$SU(3)$ orthogonal" to ${\omega}^a$. The
action for QCD (7) leads to magnetic confinement of gluons as
in the London theory of Meissner effect. The dual version
exists and produces the electric confinement. The mechanism
has been extended to include quarks. Here, the magnetic and
electric confinement of quarks is obtained in addition to an
interaction between the monopole current and the Dirac
current.  

\vspace{1.0cm}

\noindent{\bf{Acknowledgements}}

\vspace{0.5cm}

I am thankful to H.Sharatchandra, Ramesh Anishetty, R.Sridhar
and G.Baskaran for very useful discussions.

\vspace{1.0cm}

\noindent{\bf{References}}

\vspace{0.5cm}

\begin{enumerate}

\item  G.'tHooft, Nucl.Phys. {\bf B190},455 (1981). 

\item  S.Mandelstam, Phys.Rev. {\bf D19},2391 (1978).

\item T.Greenshaw, H1 Colloboration and A.Doyle, ZEUS
colloboration in the  
Proceedings of the Sixth
International Workshop in DIS and QCD, April 1998; Brussels,
Belgium  

\item G.'tHooft, {\it{in High Energy Physics Proceedings}},
1975. Edited by A.Zichichi. \\
      Nucl.Phys. {\bf B138},1 (1978): {\bf B153}, 141 (1979).

\item Y.Nambu, Phys.Rev. {\bf D10}, 4262 (1974); Phys.Rep.
{\bf C23}, 250 (1975). 

\item K.-I.Kondo, Phys.Rev. {\bf D58},105016,105019 (1998).

\item L.Faddeev and A.J.Niemi, hep-th/9807069; hep-th/9812090.

\item E.Corrigan and D.Olive, Nucl.Phys. {\bf B110}, 237
(1976); \\
      E.Corrigan, D.Olive, D.B.Fairlie and J.Nuyts, Nucl.Phys.
{\bf B106}, 475 (1976). 

\item Y.M.Cho, Phys.Rev. {\bf D21},1080 (1980); {\bf D23},2415
(1981); \\
      Phys.Rev.Lett. {\bf 44},1115 (1980).

\item A.C.T.Wu and T.T.Wu, J.Math.Phys. {\bf 15}, 53 (1974).

\item W.J.Marciano and Pagels, Phys.Rev. {\bf D12}, 1093
(1975); R.Parthasarathy and K.S.Viswanathan, Phys.Lett. {\bf
B114}, 436 (1982).

\item To show that (3) solves (1), substitute (3) in (1) and
use (2). Then, $D^{ab}_{\mu}\ {\omega}^b \ =\
{\partial}_{\mu}\ {\omega}^a \ +\ f^{abc}f^{edc}\ {\omega}^b\
{\omega}^e\ {\partial}_{\mu}\ {\omega}^d.$  
We make use of the
standard relations between $f$' and $d$' tensors for $SU(3)$
group. See equations 2.8, 2.9 and 2.10 of A.J.Macfarlane,
A.Sudberg and P.H.Weise, Commun.Math.Phys. {\bf 11}, 77
(1968).Then, $f^{abc}f^{edc}\ =\
\frac{2}{3}({\delta}_{ae}{\delta}_{bd}-{\delta}_{ad}{\delta}_
{be}) + d^{aec}d^{dbc}-d^{adc}d^{bec}$. Also,  ${\omega}^b\
{\partial}_{\mu}\ {\omega}^b \ =\ 0$ using the first relation
in (2). These give $D^{ab}_{\mu}\ {\omega}^b \ =\ {\partial}_{\mu}
{\omega}^a\ +\ \frac{4}{3}\ \{ -\frac{2}{3}({\partial}_{\mu}
{\omega}^a) \ +\ d^{aec}d^{dbc} \ {\omega}^b\ {\omega}^e \ (
{\partial}_{\mu}{\omega}^d) \ -\ d^{adc}d^{bec}\ {\omega}^b\ 
{\omega}^e\ ({\partial}_{\mu}{\omega}^d)\}$. From the second
relation in (2), it follows $d^{abc}\ ({\partial}_{\mu}
{\omega}^b)\ {\omega}^c\ =\ \frac{1}{2\sqrt{3}}\ {\partial}_
{\mu} {\omega}^a$. With this, we have $d^{aec}d^{dbc}\ 
{\omega}^b \ {\omega}^e\ ({\partial}_{\mu}{\omega}^d)\ =\ 
\frac{1}{12}({\partial}_{\mu}{\omega}^a)$ and $d^{adc}d^{bec}
\ {\omega}^b\ {\omega}^e\ ({\partial}_{\mu}{\omega}^d)\ =\ 
\frac{1}{6}\ ({\partial}_{\mu}{\omega}^a)$. Then, $D^{ab}_
{\mu}\ {\omega}^a\ =\ {\partial}_{\mu}{\omega}^a\ +\ \frac
{4}{3}\{-\frac{2}{3}+\frac{1}{12}-\frac{1}{6}\}\ ({\partial}
_{\mu}{\omega}^a)\ =\ 0$.   

\item P.Roman, {\it{Advanced Quantum Theory - \\
      An outline of the fundamental ideas}} \\
      Addison-Wesley Pub.Co.Inc. Reading, Massachusetts; 1965.

\item V.N.Gribov, Nucl.Phys. {\bf B139}, 1 (1978). 

\item From (3), we have
${\partial}_{\mu}A^a_{\nu}-{\partial}_{\nu}A^a_{\mu}\ =\ (
{\partial}_{\mu}C_{\nu}-{\partial}_{\nu}C_{\mu}){\omega}^a
+C_{\nu}({\partial}_{\mu}{\omega}^a)-C_{\mu}({\partial}_{\nu}
{\omega}^a)-\frac{8}{3g}f^{abc}{\partial}_{\mu}{\omega}^b \ 
{\partial}_{\nu}{\omega}^c$. Using (4), the non-Abelian part
is $gf^{abc}A^b_{\mu}A^c_{\nu}\ =\ C_{\mu}({\partial}_{\nu}
{\omega}^a)-C_{\nu}({\partial}_{\mu}{\omega}^a) + \frac{16}
{9g}f^{abc}f^{bed}f^{c\ell m}{\omega}^e{\omega}^{\ell}(
{\partial}_{\mu}{\omega}^d)({\partial}_{\nu}{\omega}^m)$, so
that $F^a_{\mu\nu}\ =\ ({\partial}_{\mu}C_{\nu}-{\partial}_
{\nu}C_{\mu}){\omega}^a\ -\ \frac{8}{3g}f^{abc}({\partial}_
{\mu}{\omega}^b)({\partial}_{\nu}{\omega}^c)\ +\ \frac{16}
{9g}f^{abc}f^{bed}f^{c\ell m}{\omega}^e{\omega}^{\ell}
{\partial}_{\mu}{\omega}^d\ {\partial}_{\nu}{\omega}^m$.  
Use of Jacobi identity and the relation (4) for the last term
in the previous line (the one involving three $f$'s) yields,
$F^a_{\mu\nu}\ =\ ({\partial}_{\mu}C_{\nu}-{\partial}_{\nu}
C_{\mu}){\omega}^a\ -\ \frac{4}{3g}f^{abc}{\partial}_{\mu}
{\omega}^b\ {\partial}_{\nu}{\omega}^c\ -\ \frac{16}{9g}
f^{a\ell c}f^{bed}f^{mbc}{\omega}^e{\omega}^{\ell}
{\partial}_{\mu}{\omega}^d\ {\partial}_{\nu}{\omega}^m$. Now
using $f^{edb}f^{cmb}\ =\ \frac{2}{3}({\delta}_{ec}{\delta}
_{dm}\ -\ {\delta}_{em}{\delta}_{cd})\ +\ d^{ecb}d^{dmb}\ -\
d^{deb}d^{emb}$ and the relations in (2), we find the last
term in the previous line vanishes. 

\item To see this, the first term in (5) trivially satisfies
(6). The second term in (5), when substituted in (6) gives
$f^{abc}f^{ced}{\omega}^b{\partial}_{\mu}{\omega}^e\ 
{\partial}_{\nu}{\omega}^d\ =\ \frac{2}{3}({\delta}_{ae}
{\delta}_{bd}\ -\ {\delta}_{ad}{\delta}_{be}){\omega}^b
{\partial}_{\mu}{\omega}^e{\partial}_{\nu}{\omega}^d\ +\ 
(d^{aec}d^{bdc}\ -\ d^{bec}d^{adc}){\omega}^b\ {\partial}
_{\mu}{\omega}^e\ {\partial}_{\nu}{\omega}^d$. The Kronecker
$\delta$-terms vanish due to the first relation in (2). Use of
the second relation in (2) implies, $d^{abc}{\omega}^c\
{\partial}_{\mu}{\omega}^b\ =\ \frac{1}{2\sqrt{3}}{\partial}
_{\mu}{\omega}^a$ and this simplifies the remaining terms as 
$\frac{1}{2\sqrt{3}}d^{aec}{\partial}_{\mu}{\omega}^e\
{\partial}_{\nu}{\omega}^c\ -\
\frac{1}{2\sqrt{3}}d^{adc}{\partial}_{\mu}{\omega}^c{\partial}
_{\nu}{\omega}^d$ which cancel each other. 

\end{enumerate}  
\end{document}